\newif\ifconfver
\confverfalse      

\ifconfver
     \documentclass[10pt,twocolumn,twoside]{IEEEtran}
\else
    \documentclass[11pt,draftcls,onecolumn]{IEEEtran}
\fi

\usepackage{graphicx,booktabs,color}  
\usepackage{amsmath,epsfig,amsfonts,amssymb,graphics,psfrag,theorem,calc,subfigure,url,bm,cite}
\usepackage{float}
\usepackage{stfloats}  
\usepackage{algorithmic,algorithm}

\newlength{\twidth}
\ifconfver
\else
\fi

    \makeatletter
    \def\multilimits@{\bgroup
  \Let@
  \restore@math@cr
  \default@tag
 \baselineskip\fontdimen10 \scriptfont\tw@
 \advance\baselineskip\fontdimen12 \scriptfont\tw@
 \lineskip\thr@@\fontdimen8 \scriptfont\thr@@
 \lineskiplimit\lineskip
 \vbox\bgroup\ialign\bgroup\hfil$\m@th\scriptstyle{##}$\hfil\crcr}
    \def\Sb{_\multilimits@}
    \def\endSb{\crcr\egroup\egroup\egroup}
\makeatother

\newtheorem{Lemma}{Lemma}

\theorembodyfont{\rmfamily}

\newcommand{\comment}[1]{}

\begin{document}

\title{
A Robust Design for MISO Physical-Layer Multicasting over Line-of-Sight Channels
}

%

\author{Man-Chung Yue,  \emph{Student Member, IEEE}, \\Sissi Xiaoxiao Wu, \emph{Member, IEEE}, 
Anthony Man-Cho So, \emph{Member, IEEE} 
\thanks{S. X. Wu is the corresponding author. She, M.-C. Yue and A. M.-C. So are with the Department of Systems Engineering and Engineering
Management, The Chinese University of Hong Kong, Shatin, Hong Kong
S.A.R., China. E-mail: \{xxwu, mcyue, manchoso\}@se.cuhk.edu.hk}
}

\maketitle


\begin{abstract}
This paper studies a robust design problem for far-field line-of-sight (LOS) channels where phase errors are present. 
Compared with the commonly used additive error model, the phase error model is more suitable for capturing the uncertainty 
in an LOS channel, as the dominant source of uncertainty lies in the phase. We consider a multiple-input single-output 
(MISO) multicast scenario, in which our goal is to design a beamformer that minimizes the transmit power while satisfying 
probabilistic signal-to-noise ratio (SNR) constraints. 
The probabilistic constraints give rise to a new computational challenge, as they involve random trigonometric forms.
In this work, we propose to first approximate the random trigonometric form by its second-order Taylor expansion and
then tackle the resulting random quadratic form using a 
Bernstein-type inequality. The advantage of such an approach
is that an approximately optimal beamformer can be obtained using the standard semidefinite relaxation technique. 
In the simulations, we first show that if a non-robust design (i.e., one that does not take phase errors into account) is used, then the whole system may collapse.
We then show that our proposed method is less conservative than the existing robust design based on Gaussian 
approximation and thus requires a lower power budget.
\end{abstract}


\section{Introduction}
In wireless communication systems, a channel is called line-of-sight (LOS) if a direct link between the transmitter and the receiver is always present. Information delivery over LOS channels is an important kind of data transmission in modern wireless systems, with numerous applications including satellite communications, indoor communications, and near-base-station communications. A typical far-field multiple-input single-output (MISO) LOS channel \cite[Chapter 7.22]{tse2005fundamentals} takes the form ${\bm h} = a \cdot {\bm e}$ with ${\bm e} = \exp\left( -i2\pi d/\lambda_c  \right)\left[1, \exp(i2\pi \theta t),...,\exp(i2\pi (n-1)\theta t) \right]^T$, where $\lambda_c$ is the carrier wavelength, $d$ is the distance between the transmitter and the receiver,  $\theta$ is the normalized receiver antenna separation (normalized to the
unit of the carrier wavelength), and $a$ is the attenuation of the path. \comment{Physically, ${\bm h}$ denotes the signal direction or the spatial signature induced on the receiver antenna array by the
transmitted signal.} The magnitude of the LOS channel ${\bm h}$ is determined solely by the path attenuation $a$, which usually varies slowly and is easy to estimate in practice \cite{gharanjik2015robust, el2009collaborative, see2004direction}.
By contrast, the phase ${\bm e}$ depends on many factors such as distance, antenna position, and oscillator offsets.\footnote{The LOS channel with reflected paths can be similarly modeled as a sum of ${\bm h}$'s. The design of such channels is beyond the scope of this work.} In real-world systems, errors in those factors may originate from long distance transmission delay \cite{zheng2012generic,christopoulos2012linear} (e.g., for satellite channels), asynchronous carrier frequency (e.g., cheap oscillators in mobile terminals), or arrival delay at different antennas. Therefore, phase error has a more dominant effect in LOS channels. Although there is a vast literature on robust beamforming for LOS channels, to our best knowledge, not much has been done to incorporate phase error in the noise model. Two related works we are aware of are those by El-Keyi and Champagne \cite{el2009collaborative} and Gharanjik et al \cite{gharanjik2015robust}. In \cite{el2009collaborative}, the authors study collaborative uplink beamforming for LOS channels with phase errors and model them as ${\bm e}+{\bm \Delta}$ with ${\bm \Delta}$ being the uncertainty. This model is additive in nature (with respect to ${\bm e}$) and the channel magnitude is subject to change. Recently, the authors of \cite{gharanjik2015robust} propose modeling the phase error by adding a Gaussian noise directly to the phases of the entries of ${\bm e}$. Such a multiplicative noise model is attractive from a modeling perspective. However, this model results in a substantially more difficult design problem. 

\comment{As a consequence, when accounting for channel uncertainty for LOS, a critical error usually appears in phase, which brings us a lot of difficulties in perusing robust design approaches. In the literature, there is few work addresses this issue, to our best knowledge. Prior work in \cite{el2009collaborative, see2004direction} attempts to model an additive error on ${\bm e}$, which is bounded by a ball with fixed radius. However, the resulting corrupted channel may not preserve the same form of ${\bf h}$ and thus loses the physical meaning of LOS. Recently, the authors in \cite{gharanjik2015robust} assumed an additive error in the phase of ${\bm e}$, which is more practical from a modeling perspective. However,  the additive phase error makes the problem more difficult to deal with. This actually motivates our work in this paper. }

In this paper, we are interested in the robust beamforming design for the MISO downlink multicast LOS channel under the same phase error model as in \cite{gharanjik2015robust} (the applicability of our approach is actually not limited to MISO multicasting). Specifically, we consider the power minimization problem subject to probabilistic outage constraints. Such constraints involve random trigonometric forms, which have rarely been addressed in the beamforming literature. To obtain more tractable approximations of the probabilistic constraints, we propose to approximate the random trigonometric form by a second-order Taylor expansion and then apply the Bernstein-type inequality approach in \cite{bechar2009bernstein, wang2014outage}. The resulting formulation can then be tackled by the semidefinite relaxation (SDR) technique, thereby leading to an approximately optimal beamformer. To demonstrate the effectiveness of our approach, we compare our method with that in \cite{gharanjik2015robust} by simulations. The results show that the proposed design approach is less conservative and require less transmit power than the Gaussian approximation (GA) approach in \cite{gharanjik2015robust}.

\section{System Model and Problem Formulation}
We consider the physical-layer multicasting channel, where a base station with $n$ antennas intends to transmit a common signal to $m$ single-antenna receivers.
Suppose that the channel is quasi-static and we consider the transmit design for each code block. As such, we shall omit the time index from the notations henceforth. For $j=1,\dots,m$, let ${\bm h}_j\in\mathbb{C}^n$ denote the estimated channel for user $j$ and we focus on the problem of beamforming design that is robust against the phase error of the channel. Specifically,
we model the phase error of the $j$-th channel by a Gaussian random vector ${\bm \theta}_j\sim \mathcal{N}(0,\sigma_i^2 I)$. Denoting ${\bm e}_j=(e^{i\theta_{j1}},\dots,e^{i\theta_{jn}})^T$, the corrupted channel for user $j$ can be expressed as
\begin{equation}
\hat{{\bm h}}_j={\bm h}_j\odot {\bm e}_j,
\end{equation}
where $\odot$ is the entry-wise product.

Let $s\in\mathbb{C}$ be the common unit-power signal intended for all users. Before transmission, the signal will be precoded by a precoding vector ${\bm w}\in\mathbb{C}^n$ and the resulting signal for transmission is ${\bm w}s$. After passing through the channels, the signal obtained at the $j$-th receiver is 
\begin{equation}
g_j=\hat{{\bm h}}_j^H{\bm w}s+\epsilon_j, \quad j=1,...,m,
\end{equation}
where $\epsilon_j$ is the additive noise at user $j$ and is assumed to be Gaussian distributed with mean zero and variance $\eta_j^2$. Also, the random variables/vectors $\epsilon_1,\dots,\epsilon_m,{\bm \theta}_1,\dots,{\bm \theta}_m$ are assumed to be independent. 

As usual, the quality-of-service (QoS) at user $j$ is measured by the SNR:
\begin{equation}
\text{SNR}_j({\bm w}{\bm w}^H) = {{\bm e}_j^H{M}_j({\bm w}{\bm w}^H){\bm e}_j/\eta_j^2},
\end{equation}
where ${M}_j$ is the operator on the space of Hermitian matrices given by 
\begin{equation}
{M}_j({\bm W})={\bm W}\odot({\bm h}_j{\bm h}_j^H)^T. 
\end{equation}
We say that user $j$ is in outage if $\text{SNR}_j< \gamma$, where $\gamma >0$ is a prescribed threshold. A natural formulation of the beamforming design problem is to minimize the power while maintaining a low outage probability:
\begin{equation}\label{CCPM_vector} 
\begin{array}{c@{\quad}l}
\displaystyle\min_{{\bf w}} & \|{\bm w}\|_2^2 \\ 
\noalign{\smallskip}
\mbox{s.t.} & \Pr(\text{SNR}_j({\bm w}{\bm w}^H)>\gamma)\geq 1-\rho_j,\ j=1,\dots,m.
\end{array}
\end{equation}
Here, $\rho_j$'s are the outage probabilities. Assuming $\eta_j^2=1, \forall j$ and letting ${\bm W}={\bm w}{\bm w}^H$, we can equivalently write \eqref{CCPM_vector} as
\begin{align}\notag
({\rm P})\quad&\displaystyle\min_{{\bf W}} \quad {\rm Tr}({\bm W}) \\ \label{pcon}
\mbox{s.t.} \quad& \Pr({\bm e}_j^H{M}_j({\bm W}){\bm e}_j>\gamma)\geq 1-\rho_j,\ j=1,\dots,m,\\\label{r1}
                  &\text{Rank}({\bm W})=1,\quad {\bm W}\succeq 0.
\end{align}
Recall that we have ${\bm e}_j=(e^{i\theta_{j1}},\dots,e^{i\theta_{jn}})^T$ and ${\bm \theta}_j\sim \mathcal{N}(0,\sigma_i^2 I)$. 
Problem (P) is difficult to handle since we do not have a good analytic description of the probabilistic constraint \eqref{pcon}. In \cite{gharanjik2015robust}, the authors work around this by approximating ${\bm e}_j^H{M}_j({\bm W}){\bm e}_j$ by a Gaussian random variable with matching mean and covariance. The aim of this paper is to derive a more accurate and efficient alternative to handle \eqref{pcon}.

\section{Approximating the probabilistic constraints}
In this section, we use a Taylor series approximation and the Bernstein-type inequality to tackle \eqref{pcon}. Specifically, by considering the second-order Taylor expansion of the random variables ${\bm e}_j^H{M}_j({\bm W}){\bm e}_j$, we transform the outage probability into the large deviation of a quadratic function in Gaussian random variables. The probabilistic constraint can then be handled by the Bernstein-type inequality approach \cite{bechar2009bernstein, wang2014outage}. Finally, by employing the SDR technique, we obtain a tractable approximation of the probabilistic constraint \eqref{pcon}.

Let us first introduce some notations. Denote by $\mathcal{S}^n$ and $\mathcal{K}^n$ the sets of $n\times n$ real symmetric and skew-symmetric matrices respectively. Given any arbitrary matrix ${\bm A}$ with its $(i,j)$-th element denoted by $A_{ij}$, let $L:\mathbb{R}^{n\times n}\rightarrow\mathbb{R}^{n\times n}$ be the linear map given by
$$  \left(L({\bm A})\right)_{kl} = \left\{
\begin{array}{c@{\quad}l}
\displaystyle{A_{kk}-\sum_{j}A_{kj}} & {~\rm for~} k=l, \\
A_{kl} & {~\rm for~} k\neq l
\end{array}
\right.$$
and $f:\mathbb{R}^{n\times n}\rightarrow\mathbb{R}^n$ be the linear map given by
$$\left(f({\bm A})\right)_k = 2\sum_{j}A_{kj}.$$
To build our approximation, we need the following two lemmas, whose proofs are relegated to the appendix.
\begin{Lemma}\label{quadratic_appr}
Let ${\bm \theta}\in\mathbb{R}^n$, ${\bm x}=[e^{i\theta_1}, e^{i\theta_2},...,e^{i\theta_n}]^T$, and ${\bm M}$ be an $n\times n$ Hermitian matrix with symmetric real part ${\bm A}\in\mathcal{S}^n$ and skew-symmetric imaginary part ${\bm B}\in\mathcal{K}^n$. Then, the second-order Taylor approximation of $\bm{x}^H\bm{M}\bm{x}$ is given by
\begin{equation}
{\bm x}^H{\bm M}{\bm x}\approx \sum_{k,l}M_{kl}+{\bm \theta}^TL({\bm A}){\bm \theta}+f({\bm B})^T{\bm \theta}.
\end{equation}
\end{Lemma}
Lemma \ref{quadratic_appr} allows us to approximate the hard probabilistic constraint \eqref{pcon} by a more well-studied one. In particular, if ${\bm \theta}\sim \mathcal{N}(0,\sigma^2{\bm I})$ and $\sigma$ is small, then we should have
\begin{align}\label{appr_eq}
&\Pr\!\left({\bm x}^H{\bm M}{\bm x}\geq \gamma\right)\!\notag\\
\approx&\!\Pr\!\left(\sum_{k,l}M_{kl}+\sigma^2 {\bm \xi}^TL({\bm A}){\bm \xi}+\sigma f({\bm B})^T{\bm \xi}\geq \gamma\right),
\end{align}
where ${\bm \xi}\sim \mathcal{N}(0,{\bm I})$ is the standard $n$-dimensional Gaussian random vector. Given the above approximation, it suffices to approximate the right hand side of \eqref{appr_eq}, which can be achieved using the following lemma:
\begin{Lemma}\label{cc_to_SDP}
Let $({\bm A},{\bm B},{\bm Q},{\bm b},y,z)$ be a solution to the system
\begin{equation}\label{system_S}\tag{$S_{\gamma,\rho}$}
\begin{array}{c@{\quad}l}
\noalign{\smallskip}
 &\hspace{-5mm} \left\{
\begin{array}{c@{\quad}l}
\displaystyle{\sum_{k,l}A_{kl}\!+\!{\rm Tr}({\bm Q})\!-\!2z\sqrt{\log\rho^{-1}}\!+\!2y\log\rho\!\geq\! \gamma}, \\
\sqrt{\|{\bm Q}\|_F^2+\frac{1}{2}\|{\bm b}\|_2^2}\leq z,\\
y{\bm I}+{\bm Q}\succeq \mathbf{0},\\
{\bm Q}=\sigma^2L({\bm A}),\\
{\bm b}=\sigma f({\bm B}),\\
y\geq 0,\\
{\bm Q}, {\bm A}\in\mathcal{S}^n, {\bm B}\in\mathcal{K}^n.
\end{array}
\right.
\end{array}
\end{equation}
Then, ${\bm M}={\bm A}+i{\bm B}$ satisfies
\begin{equation}\label{quad_cc}
\Pr\left(\sum_{k,l}M_{kl}+\sigma^2 {\bm \xi}^TL({\bm A}){\bm \xi}+\sigma f({\bm B})^T{\bm \xi}\geq \gamma\right)\geq 1-\rho.
\end{equation}
\end{Lemma}
With a slight abuse of notation, we write $({\bm A},{\bm B})\in$ \ref{system_S} if there exist ${\bm Q}$, ${\bm b}$, $y$ and $z$ such that $({\bm A},{\bm B},{\bm Q},{\bm b},y,z)$ is feasible to the system \eqref{system_S}. Also, let $A_j({\bm W})$ and $B_j({\bm W})$ be the real and imaginary parts of $M_j(\bm{W})$, respectively. Using Lemmas \ref{quadratic_appr} and \ref{cc_to_SDP}, we obtain the following approximation of Problem (P):
\begin{align}
({\rm AP})\quad &\displaystyle\min_{{\bm W}} \quad{\rm Tr}({\bm W} ) \notag\\ \label{cont}
&\mbox{s.t.} \quad(A_j({\bm W}),B_j({\bm W}))\in S_{\gamma,\rho},\ j=1,\dots,m,\\\notag
             & \quad\quad\text{Rank}({\bm W})=1,\quad {\bm W}\succeq 0.
\end{align}
By dropping the non-convex rank-one constraint, we arrive at
\begin{align*}
({\rm APR})\quad \displaystyle\min_{{\bm W}\succeq 0} & \quad{\rm Tr}({\bm W} ) \quad\quad \mbox{s.t.} \quad \eqref{cont}~{\rm satisfied}.
\end{align*}

Problem (APR) is an instance of semidefinite programming (SDP) and hence can be solved efficiently by any off-the-shelf SDP solver \cite{grant2008cvx}. The optimal solution ${\bm W}^\star$ to Problem (APR) can be of any rank because we drop the rank constraint. For ease of presentation, we consider the problem instances where Problems (P), (AP), and (APR) are feasible. Then, to extract a rank-one solution from $\bm{W}^\star$, we can use Gaussian rounding (see Algorithm \ref{alg}) to obtain a vector beamformer $\hat{\bm{w}}$ such that $\hat{\bm{w}}\hat{\bm{w}}^H$ is approximately feasible to Problem (P). We remark that although Problem (APR) is convex and tractable, its feasible region is not necessarily included in that of (P) and hence theoretically the beamformer $\hat{\bm{w}}$ can fail to satisfy the original outage probability requirement. However, our numerical results in the next section show that the proposed approximation is actually \emph{empirically safe} under our scenario settings: all the outage probabilities given by $\hat{\bm{w}}$ are always smaller than $\rho_j$'s. 
\begin{algorithm}[H]
\caption{Gaussian Randomized Rounding} \label{alg}
\begin{algorithmic}[1]
\STATE Input: Optimal solution ${\bm W}^\star$ to (APR), number of trials $I$.
\IF {${\rm Rank}({\bm W}^\star)>1$}
\FOR {$i=1,2,\ldots,I$}
\STATE Generate $\tilde{\bm w}_i\sim \mathcal{CN}({\bf 0}, {\bm W}^\star)$.
\STATE Scale $\tilde{\bm w}_i$ so that $(A_j(\tilde{\bm w}_i\tilde{\bm w}_i^H),B_j(\tilde{\bm w}_i\tilde{\bm w}_i^H))\in S_{\gamma,\rho}$ for all $j=1,\dots,m$.
\ENDFOR
\STATE Set $\hat{\bm w}={\rm argmin}_i {\rm Tr}(\tilde{\bm w}_i\tilde{\bm w}_i^H)$.
\ELSE
\STATE Set $\hat{\bm w}$ such that $\hat{\bm w}\hat{\bm w}^H={\bm W}^\star$.
\ENDIF
\STATE Output: A beamformer $\hat{\bm w}$ with $\hat{\bm w}\hat{\bm w}^H$ feasible to (AP).
\end{algorithmic}
\end{algorithm}

\section{numerical simulations}
In this section, we provide numerical simulations to demonstrate the superiority of the proposed robust design. The setup of the experiment is as follows: the number of transmit antennas is $n=8$; the number of users served is $m=16$; channels are generated by ${\bm h}_i \sim \mathcal{CN}({\bf 0}, {\bf I})$ independently; the noise power at each user is set to be $1$; the phase error variance is $\sigma_i^2=5/360, \forall i$; the SNR outage probability is set to be ${\rho_j=0.1}, \forall j$. We averaged $100$ channel realizations to get the plots. 

\begin{figure}[t]
\centering
\includegraphics[width = 9cm]{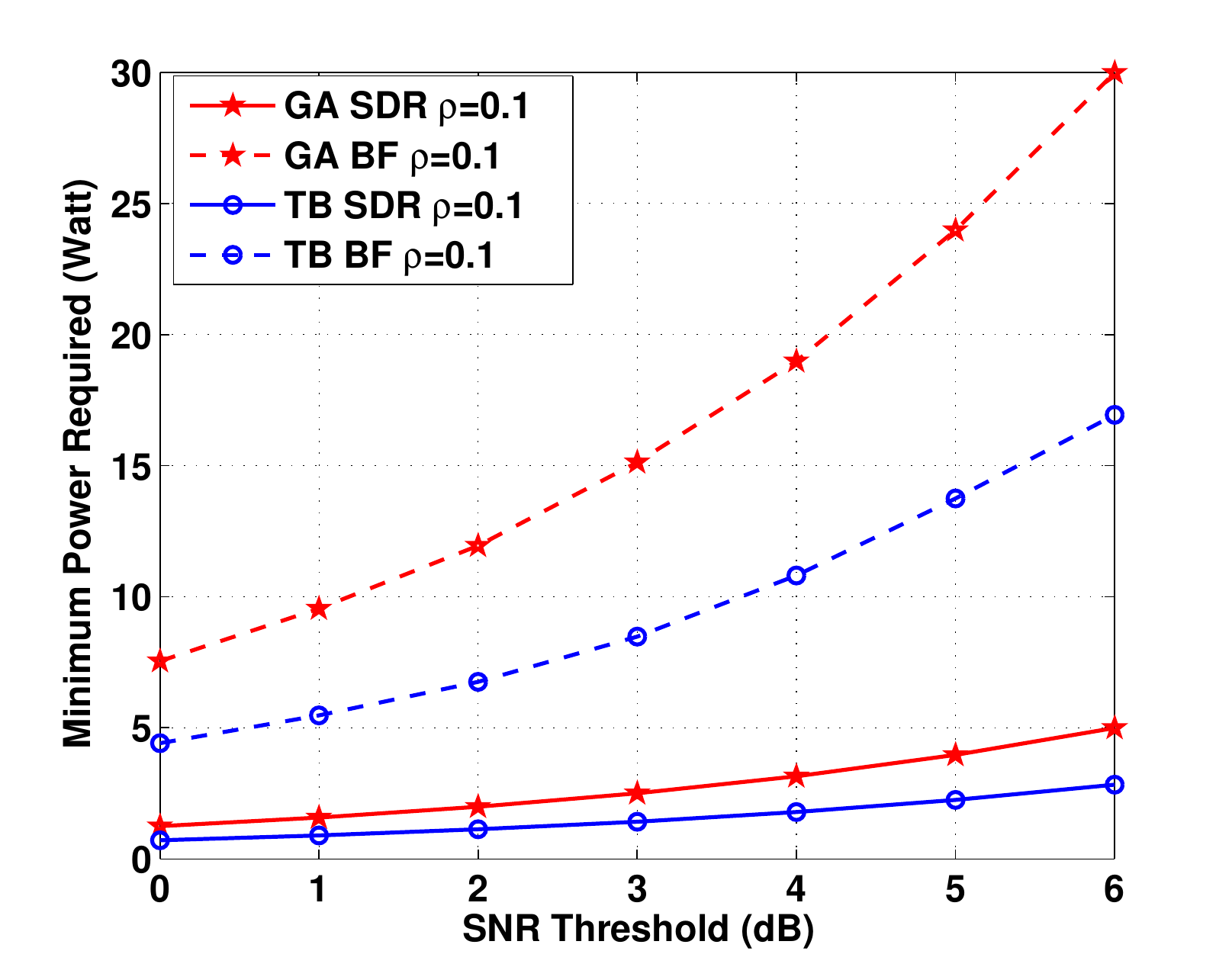}
\vspace{-0.2cm}
\caption{{\protect\footnotesize {Minimum power required versus the SNR threshold when the SNR satisfaction probability is equal to $\rho = 0.1$.}}}
\label{Fig:1}
\end{figure}

\begin{figure}[t]
\centering
\includegraphics[width = 9cm]{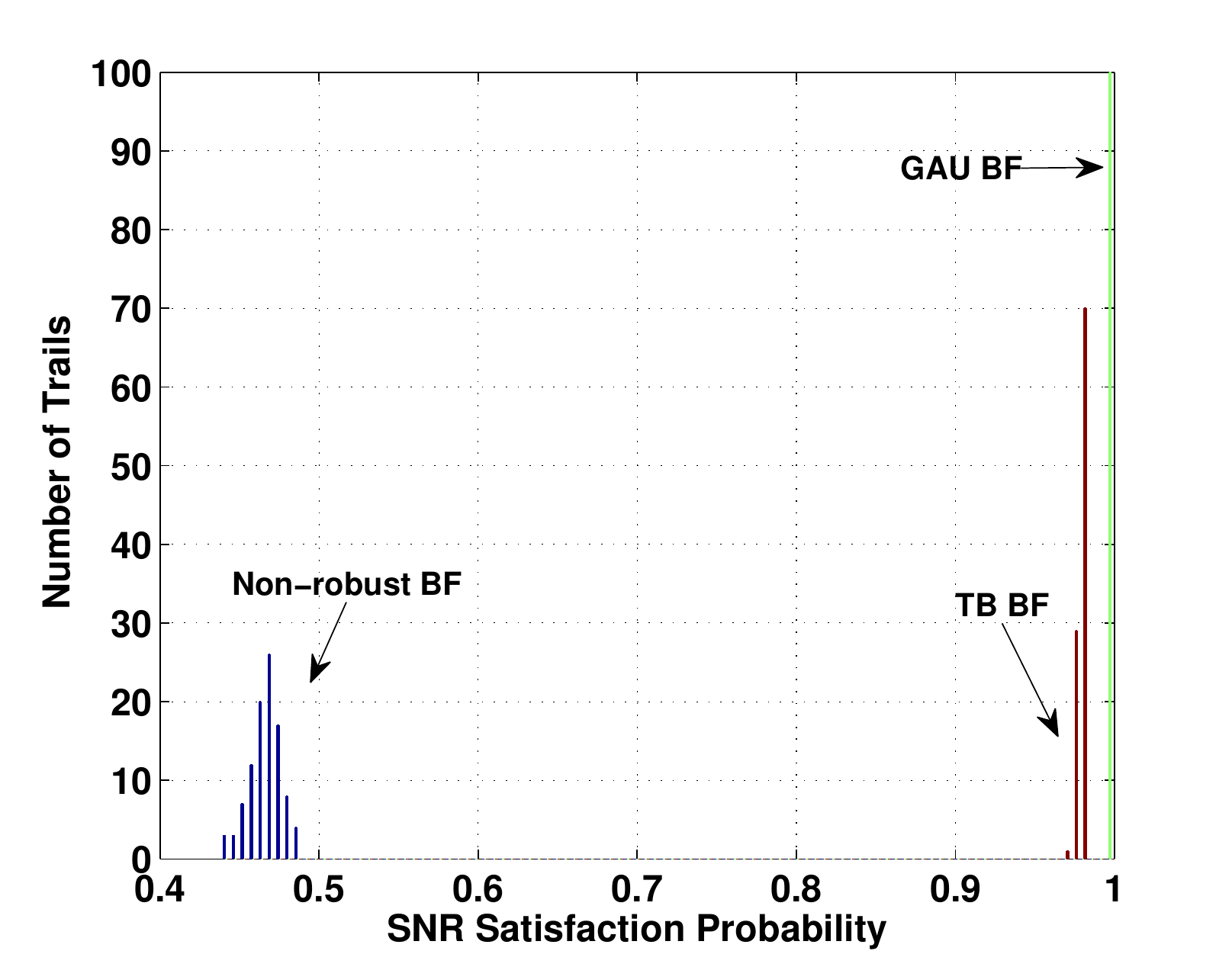}
\vspace{-0.2cm}
\caption{{\protect\footnotesize {Histogram of the SNR satisfaction probability.}}}
\label{Fig:2}
\end{figure}

In Figure \ref{Fig:1}, we show the minimum power required to satisfy the outage constraints as the SNR threshold varies from $0$dB to $6$dB when $\rho = 0.1$. Specifically, we compare the GA design in \cite{gharanjik2015robust} with the proposed design (called TB in the legend). From the plots, we see that the optimal values of the SDRs serve as lower bounds of the respective beamforming schemes. In the legend, ``SDR'' refers to the values obtained by the SDRs prior to rounding and ``BF'' refers to the actual power required by the beamformers produced by the rounding. The figure shows that the proposed TB approach needs a lower power budget than the GA approach in \cite{gharanjik2015robust}.
To investigate the conservatism, we plot the histogram of the SNR satisfaction probability of the beamformers obtained by different schemes; see Figure \ref{Fig:2}. In the non-robust approach, the channel errors are simply ignored and the problem reduces to the classical physical-layer multicasting design problem \cite{MulticastSidiropoulos06, MainPaper}.  
The non-robust histogram reveals that if we do not take phase error into account, the outage probability requirement could be seriously violated and the system is totally unreliable. This actually demonstrates the importance of the robust designs. For the GA approach, the SNR satisfaction probability is always $1$ under our setting. This means that the GA scheme sacrifices extra transmit power for unnecessary conservatism. This is also manifested in Figure \ref{Fig:1}. For the proposed TB approach, we see that the SNR satisfaction probability is less than $1$ but exceeds the target threshold $0.9$. Hence, the proposed TB approach is reliable but less conservative than the GA counterpart. Both figures demonstrate the superiority of the proposed TB design.

\section{conclusion}
In this paper, we study the robust design for LOS channels with phase uncertainty. In particular, we consider the MISO physical-layer multicasting scenario and aim at minimizing the transmit power subject to probabilistic SNR constraints. We employ Taylor series, a Bernstein-type inequality and the SDR technique to approximate the probabilistic SNR constraints by SDPs, from which an approximate beamformer can be obtained. According to our simulation results, the robust design is important when phase error is present. Furthermore, the proposed method is less conservative and requires lower transmit power when compared with the existing design approach based on Gaussian approximations. One future direction is to extend the proposed method to study more general LOS channels with reflected paths.

\vspace{-0.1cm}
\section{Appendix}
\subsection{Proof of Lemma \ref{quadratic_appr}}
Let $\theta_{kl}=\theta_k-\theta_l$ and $\Re(\cdot)$ be the real part of a complex number/vector/matrix. We have
\vspace{-0.05cm}
\begin{equation*}
\begin{split}
{\bm x}^H{\bm M}{\bm x}&=\sum_{k,l}M_{lk}e^{i(\theta_k-\theta_l)}\\
\/
&=\sum_{k}M_{kk}+\sum_{k>l}M_{lk}e^{i\theta_{kl}}+\sum_{k>l}M_{lk}e^{i\theta_{lk}}\\
\/
&=\sum_{k}M_{kk}+2\sum_{k>l}\Re\left(M_{lk}e^{i\theta_{kl}}\right)\\
\/
&=\sum_{k}M_{kk}+2\sum_{k>l}r_{lk}\cos(\phi_{lk}+\theta_{kl}).
\end{split}
\end{equation*}
\vspace{-0.1cm}
To handle the cosine terms, we use Taylor approximation:
\begin{equation*}
\begin{split}
\cos(\phi_{lk}+\theta_{kl})&=\cos\phi_{lk}\cos\theta_{kl}-\sin\phi_{lk}\sin\theta_{kl}\\
&\approx\cos\phi_{lk}-\frac{\theta_{kl}^2}{2}\cos\phi_{lk}-\theta_{kl}\sin\phi_{lk}.
\end{split}
\end{equation*}
Thus,
\begin{equation*}
\begin{split}
{\bm x}^H{\bm M}{\bm x}\approx&\sum_{k}M_{kk}+2\sum_{k>l}r_{lk}\cos\phi_{lk}\\
&-2\sum_{k>l}r_{lk}\left(\frac{\theta_{kl}^2}{2}\cos\phi_{lk}+\theta_{kl}\sin\phi_{lk} \right)\\
=&\sum_{k}M_{kk}+\sum_{k\neq l}r_{kl}\cos\phi_{kl}\\
&-\sum_{k>l}A_{kl}\theta_{kl}^2+2\sum_{k>l}B_{kl}\theta_{kl}\ \ (\because \phi_{kl}=-\phi_{lk})\\
=&\sum_{k,l}M_{kl}-\sum_{k>l}A_{kl}\theta_{kl}^2+2\sum_{k>l}B_{kl}\theta_{kl}.
\end{split}
\end{equation*}

\noindent Since ${\bm A}\in\mathcal{S}^n$ and ${\bm B}\in\mathcal{K}^n$, we have
$
\sum_{k>l}A_{kl}\theta_{kl}^2=\frac{1}{2}\sum_{k,l}A_{kl}\theta_{kl}^2 \label{symmetric_A}
$
and
$
2\sum_{k>l}B_{kl}\theta_{kl}=\sum_{k,l}B_{kl}\theta_{kl}.\label{skew_symmetric_B}
$
Therefore, we have
\begin{equation*}\label{eq_AB}
\begin{split}
&-\sum_{k>l}A_{kl}\theta_{kl}^2+2\sum_{k>l}B_{kl}\theta_{kl}\\
\/
=&-\frac{1}{2}\sum_{k,l}A_{kl}\theta_{k}^2-\frac{1}{2}\sum_{k,l}A_{kl}\theta_{l}^2+\sum_{k,l}A_{kl}\theta_k\theta_l\\
&+\sum_{k,l}B_{kl}\theta_k-\sum_{k,l}B_{kl}\theta_l\\
=&-\!\sum_{k}\!\left(\theta_k^2\sum_{l}A_{kl}\right)\!\!+\!\sum_{k,l}A_{kl}\theta_k\theta_l\!+\!2\sum_k\left(\theta_k\sum_{l}B_{kl}\right)\\
=&\ {\bm \theta}^TL({\bm A}){\bm \theta}+f({\bm B})^T{\bm \theta},
\end{split}
\end{equation*}
which implies that 
${\bm x}^H{\bm M}{\bm x}\approx\sum_{k,l}M_{kl}+{\bm \theta}^TL({\bm A}){\bm \theta}+f({\bm B})^T{\bm \theta}$.
This completes the proof.

\vspace{-0.2cm}

\subsection{Proof of Lemma \ref{cc_to_SDP}}
By \cite[Fact 1]{wang2014outage}, we have 
\begin{equation}\label{quad_cc_2}
\begin{split}
&\Pr\!\Big(\!\sigma^2{\bm \xi}^TL({\bm A}){\bm \xi}\!+\!\sigma f({\bm B})^T{\bm \xi}\geq \sigma^2{\rm Tr}(L({\bm A}))\\
&\!-\!2\sigma^2\lambda\!^-\!(L({\bm A}))\tau\!-\!2\sqrt{\sigma^4\|L({\bm A})\|_F^2+\frac{1}{2}\sigma^2\|f({\bm B})\|_2^2}\sqrt{\tau}\Big)\\
&\geq 1-e^{-\tau},
\end{split}
\end{equation}
where $\lambda^-(L({\bm A}))=\max\{\lambda_{\max}\left(-L({\bm A})\right),0\}$. Define $\zeta:(0,\infty)\rightarrow (-\infty,\sigma^2{\rm Tr}(L({\bm A})))$ to be the function 
\begin{equation}
\begin{split}
\zeta(\tau)=&\ \sigma^2{\rm Tr}(L({\bm A}))-2\sigma^2\lambda^-(L({\bm A}))\tau\\
&-2\sqrt{\sigma^4\|L({\bm A})\|_F^2+\frac{1}{2}\sigma^2\|f({\bm B})\|_2^2}\sqrt{\tau}.
\end{split}
\end{equation}
Then, $\zeta$ is strictly decreasing and $\zeta^{-1}$ is well defined on $(-\infty,\sigma^2{\rm Tr}(L({\bm A})))$. Let $\bar{\gamma}=\gamma-\sum_{k,l}M_{kl}$. Suppose that
\[
\gamma< \sum_{k,l}A_{kl}-\sigma^2\sum_{k,l}A_{kl}+\sigma^2\sum_{k}A_{kk}.
\]
It follows that
\begin{equation*}
\bar{\gamma}=\gamma-\sum_{k,l}A_{kl}<-\sigma^2\sum_{k,l}A_{kl}+\sigma^2\sum_{k}A_{kk}=\sigma^2{\rm Tr}(L({\bm A}))
\end{equation*}
and thus \eqref{quad_cc_2} can be rewritten as 
\begin{equation}
\Pr\left(\sum_{k,l}M_{kl}+{\bm \theta}^TL({\bm A}){\bm \theta}+f({\bm B})^T{\bm \theta}\geq \gamma\right)\geq 1-e^{-\zeta^{-1}(\bar{\gamma})}.
\end{equation}
This leads to the following sufficient condition for \eqref{quad_cc}:
\begin{equation*}
1-e^{-\zeta^{-1}(\bar{\gamma})}\geq 1-\rho\Leftrightarrow \zeta^{-1}(\bar{\gamma})\geq -\log\rho\Leftrightarrow \bar{\gamma}\leq \zeta(-\log\rho).
\end{equation*}
Written out explicitly, the above becomes
\begin{equation}
\begin{split}
&\sum_{k,l}A_{kl}+\sigma^2{\rm Tr}\left( L({\bm A})\right)+2\sigma^2\lambda^-\left(L({\bm A}) \right)\log\rho\\
&-2\sqrt{-\log\rho}\sqrt{\sigma^4\|L({\bm A})\|_F^2+\frac{\sigma^2}{2}\|f({\bm B})\|_2^2}\geq \gamma.
\end{split}
\end{equation}
This constraint can equivalently be expressed as the system ($S_{\gamma,\rho}$). This completes the proof.

\newpage


\end{document}